\documentclass[twocolumn, twocolappendix]{aastex631}
\usepackage{enumitem}
\usepackage{xspace}
\usepackage{amsmath}
\usepackage{float}
\usepackage{booktabs}

\newcommand{\sw}[1]{\texttt{#1}}

\shorttitle{Measuring the magnetization profiles of galaxy groups and clusters}
\shortauthors{Sharma et al.}

\begin{document}
% \linenumbers

\title{Measuring the magnetization profiles of galaxy groups and clusters}

\correspondingauthor{Kritti Sharma}
\email{kritti@caltech.edu}

\author[0000-0002-4477-3625]{Kritti Sharma}
\affiliation{Cahill Center for Astronomy and Astrophysics, MC 249-17 California Institute of Technology, Pasadena CA 91125, USA.}

\author[0000-0002-7252-5485]{Vikram Ravi}
\affiliation{Cahill Center for Astronomy and Astrophysics, MC 249-17 California Institute of Technology, Pasadena CA 91125, USA.}
\affiliation{Owens Valley Radio Observatory, California Institute of Technology, Big Pine CA 93513, USA.}

\author[0000-0001-8356-2014]{Elisabeth Krause}
\affiliation{Department of Astronomy/Steward Observatory, University of Arizona, 933 North Cherry Avenue, Tucson, AZ 85721, USA}
\affiliation{Department of Physics, University of Arizona, 1118 E Fourth Street, Tucson, AZ 85721,
USA}

\begin{abstract}

Magnetic fields are vital in determining the physical conditions of baryons in massive halos. They are, however, notoriously difficult to observe. This impacts problems from our understanding of feedback and self-regulation in galaxy formation, to the observed deficit in the amplitude of thermal Sunyaev--Zel'dovich (tSZ) power spectrum on all scales that affects cosmological inferences. Although Faraday rotation measures (RMs) offer a direct probe of magnetic fields, the RM observable weights magnetic fields by the free-electron density. Several direct probes of free electrons, including fast radio burst (FRB) dispersion measures (DMs) and kinematic SZ (kSZ) effect, are now maturing into precision tools. We combine RM$^2$ stacks from SPICE-RACS DR2 around DESI galaxy groups and ROSAT galaxy clusters with optical depth profiles constrained jointly by the FRB DM--redshift relation and the kSZ effect, and obtain 95\% upper limits on the rms line-of-sight magnetic field of $<3.9$--$11.6~\mu$G ($<2.3$--$6.4~\mu$G) in groups (clusters) for a coherence length of 10~kpc. Combining these with thermal pressure profiles inferred from tSZ--cosmic shear cross-correlations, we obtain 95\% lower limits on the plasma $\beta$ of $>0.31$ ($>11$) in groups (clusters). Although current constraints are only conditional bounds, we forecast a first detection with larger datasets from the Deep Synoptic Array (DSA). We present the first forecast of the DSA RM grid density, conservatively finding $\sim365$ RMs\,deg$^{-2}$, which when combined with 10$^4$ FRBs would measure the field in groups to $3$--$13\%$ precision, compared to current constraints of $38$--$75\%$. Although the interpretation of current measurements remains challenging owing to depolarization along turbulent halo sightlines and the unknown coherence length, larger samples and DSA/Square Kilometer Array observations may resolve these uncertainties by measuring polarization-fraction profiles and the RM structure function using extended radio sources, respectively.

\end{abstract}

\section{Introduction} \label{sec:introduction}

Magnetic fields ubiquitously permeate every scale of cosmic structure, from the interstellar medium of galaxies to the diffuse filaments of the cosmic web, yet their origin and evolution remain among the least constrained aspects of structure formation~\citep{2004IJMPD..13.1549G, 2008RPPh...71d6901K}. A central open question is to what degree the magnetic fields observed today trace primordial seed fields generated in the very early universe~\citep{2021RPPh...84g4901V}, and/or fields generated later by astrophysical sources, through dynamo amplification in galaxies and subsequent injection into the intergalactic medium (IGM) by active galactic nucleus (AGN) outflows~\citep{2023ARA&A..61..561B}. Distinguishing between primordial magnetogenesis scenarios~\citep{2016PhRvD..94l1302M, 2012JCAP...10..034F, 2024ApJ...977..128M} requires isolating the volume-filling primordial component from the magnetized bubbles produced by feedback~\citep{2018MNRAS.480.5113M, 2022MNRAS.515.5673A}. Because any density-dependent probe of magnetic fields is necessarily weighted towards the densest, most strongly magnetized environments, properly modeling this feedback contamination is essential for any inference of the primordial fields~\citep{2025A&A...693A.208C}.

Beyond magnetogenesis, magnetic fields may play a significant role in galaxy formation and the baryon cycle. Magnetic fields regulate cosmic-ray transport across galaxies and halos, shaping feedback through the streaming and diffusion of cosmic rays injected by supernovae and AGNs~\citep{2023A&ARv..31....4R, 2016PhRvL.116s1302P}.

Magnetic fields may also help resolve a persistent puzzle in cluster and group astrophysics: the thermal Sunyaev--Zel'dovich (tSZ) auto-power spectrum is systematically lower on all scales than theoretical predictions~\citep{2014MNRAS.440.3645M, 2025MNRAS.540.1055E}. This discrepancy is commonly attributed to strong baryonic feedback, but non-thermal pressure support from turbulence, cosmic rays, and magnetic fields can also suppress the tSZ signal, and observations currently lack an independent means of disentangling these contributions~\citep{2010ApJ...725.1452S}. Measurements of halo magnetic fields can therefore constrain both the non-thermal pressure budget and the physics of feedback itself. In particular, the ratio of thermal to magnetic pressure, parameterized by $\beta \equiv P_\mathrm{th}/P_B$, remains poorly constrained observationally, leaving a key aspect of halo gas physics uncertain despite its importance for interpreting feedback~\citep{2023A&ARv..31....4R}.

Faraday rotation offers a direct observational route to these fields~\citep{1968PASJ...20..388S}. Magnetoionic media are birefringent, and therefore, as a linearly polarized electromagnetic wave of wavelength $\lambda$ propagates through a magnetized plasma, its polarization angle rotates by an amount $\Delta \Phi = \mathrm{RM} \lambda^2$, set by the Faraday rotation measure~\citep[RM;][]{1963Natur.197.1162G},
\begin{equation}
    \mathrm{RM} = 0.812 \int_0^{l} (1+z(l^\prime))^{-2} \left(\frac{n_\mathrm{e}}{\mathrm{cm^{-3}}}\right) \left(\frac{B_\parallel}{\mu\mathrm{G}}\right) \left(\frac{\mathrm{d}l^\prime}{\mathrm{pc}}\right) \ \frac{\mathrm{rad}}{\mathrm{m^{2}}},
    \label{eqn:RM}
\end{equation}
where $n_e$ is the free electron density, $B_\parallel$ is the line-of-sight magnetic field, and the redshift factor $(1+z(l'))^{-2}$ accounts for the redshifting of the wavelength at which the rotation physically occurs, at each point $l^\prime$ along the sightline. Equation~\ref{eqn:RM} makes explicit the central observational challenge: RM measures the \textit{product} of electron density and magnetic field, so a single RM measurement can never disentangle a weak field threading dense gas from a strong field threading diffuse gas. This is the baryon-magnetism degeneracy, and it is precisely this weighting towards the densest, most strongly magnetized sightline segments that ties the RM integral to the feedback-driven bubble contamination. Faraday rotation intrinsic to the polarized radio sources themselves, in their host galaxies and near-source environments, adds RM scatter that is unrelated to the foreground gas and therefore acts as noise~\citep{2010MNRAS.409L..99S, 2015A&A...575A.118O, 2025ApJ...987..173S}. 

The imminent availability of enormous polarized radio source catalogs from next-generation radio surveys, culminating in $> 10^6$ RMs with the Deep Synoptic Array~\citep[DSA;][]{2019BAAS...51g.255H} and the Square Kilometer Array~\citep[SKA;][]{2004NewAR..48..979C}, will dramatically reduce the statistical uncertainty on RM-based measurements, but does nothing to resolve the underlying degeneracy. Progress requires an independent measurement of $n_e$ along statistically matched sightlines. Dispersion measures (DMs) of fast radio bursts (FRBs) and kinematic SZ (kSZ) effect optical depth encode the line-of-sight integral of the free electron density, thus providing exactly this missing ingredient. The DSA FRB survey, together with the kSZ effect measurements from the next-generation cosmic microwave background experiments such as the Simons Observatory~\citep{2025JCAP...08..034A}, will deliver independent gas density measurements for the \emph{same} halo population probed by RM measurements, offering a direct path to breaking the degeneracy in Equation~\ref{eqn:RM}.

Another complication concerns how to stack RM measurements statistically. A naive stack of the raw signed RM around galaxies or clusters suffers severe signal loss from field reversals along the line of sight and from sightline-to-sightline. Stacking $|\mathrm{RM}|$ instead avoids cancellation but introduces a noise bias, since $\langle |\mathrm{RM} + n| \rangle \neq \langle |\mathrm{RM}| \rangle$ for any zero-mean measurement noise $n$, which suppresses the inferred $|\mathrm{RM}|\times g$ amplitude by a factor of $\sim2$ at realistic noise levels~\citep{2026JCAP...07..096Z}. Stacking $\mathrm{RM}^2$ instead avoids both pathologies: it is immune to sign cancellation, unbiased by symmetric measurement noise, and admits a clean theoretical interpretation as a bispectrum of the electron-density-weighted line-of-sight magnetic field with the tracer overdensity field~\citep{2026JCAP...07..096Z}.

\begin{figure*}[ht!]
\centering
\includegraphics[width=0.32\textwidth]{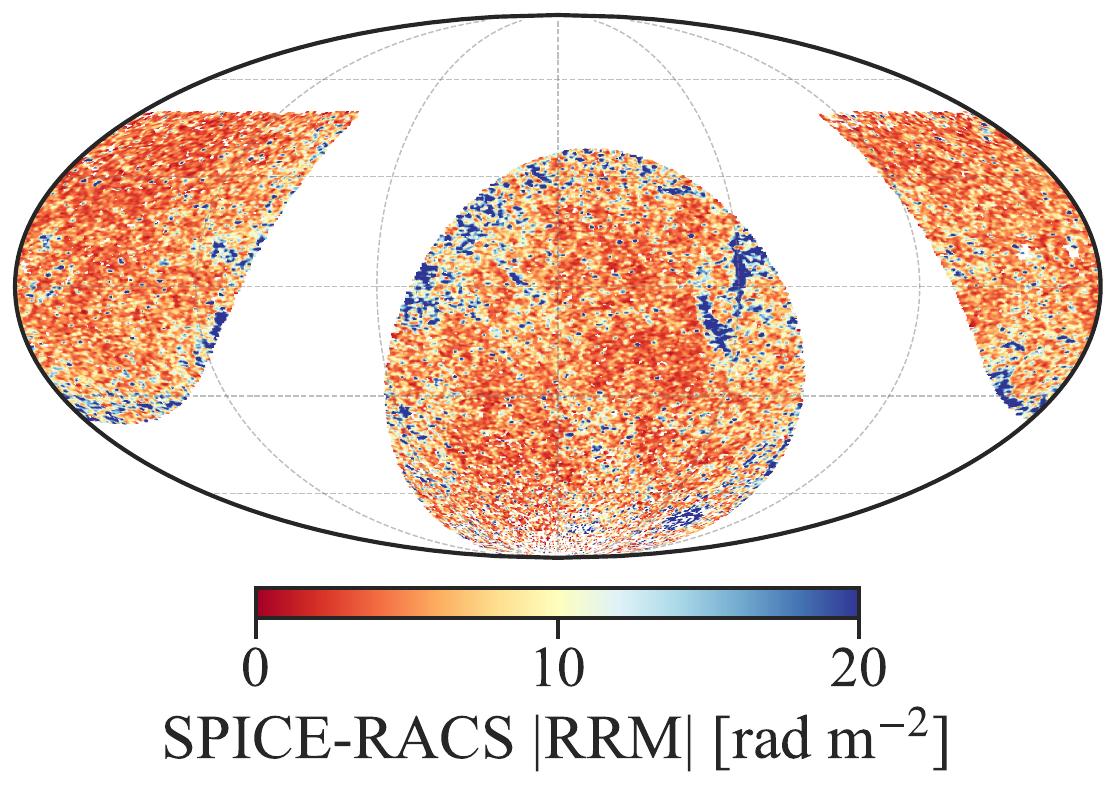}
\includegraphics[width=0.32\textwidth]{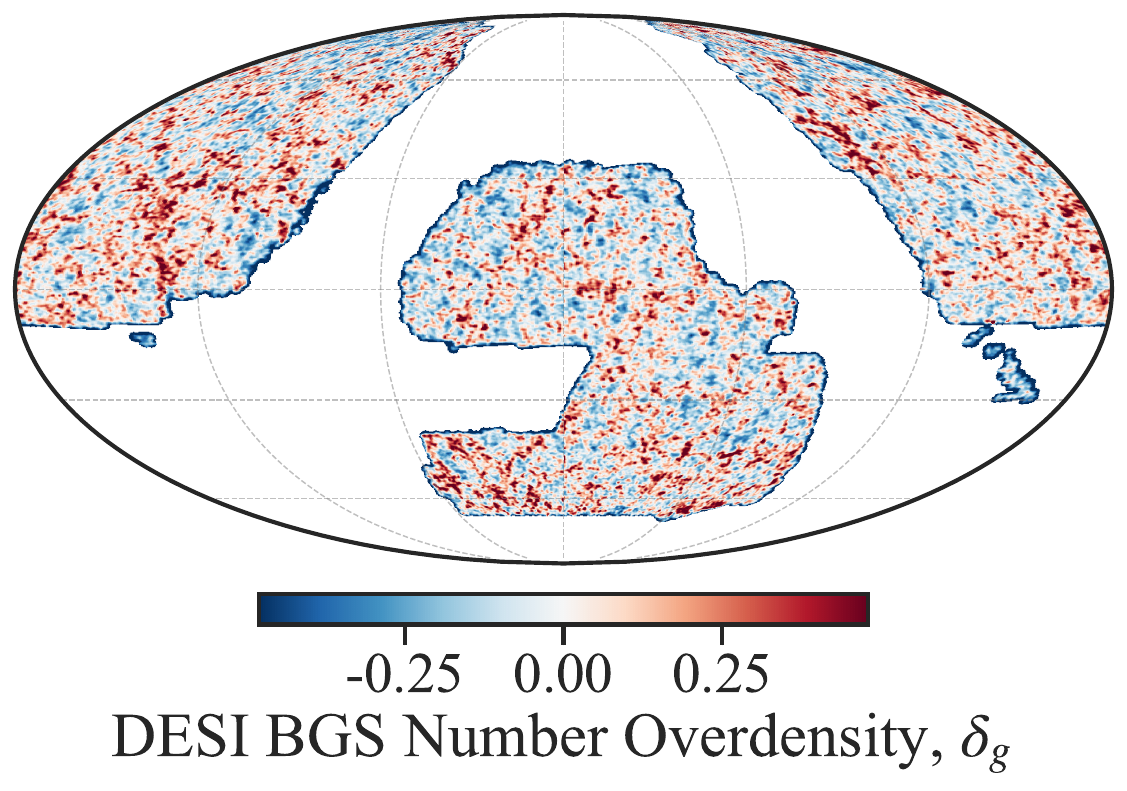}
\includegraphics[width=0.32\textwidth]{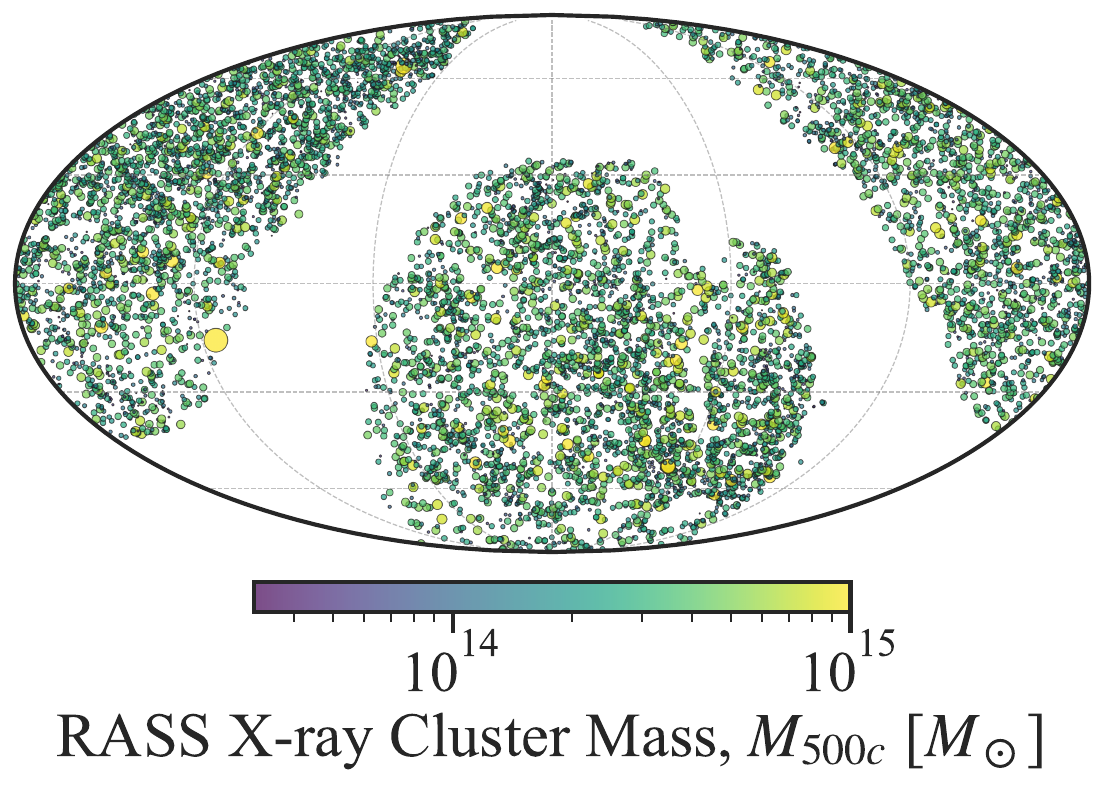}
\caption{Sky distribution of the datasets used in this work, shown in Mollweide projection. \textit{Left:} Galactic-foreground-subtracted RM amplitude, $|\mathrm{RRM}|$, from the SPICE-RACS DR2 catalog~\citep{2026PASA...43...89T}, masked at Galactic latitude $|b| < 25^\circ$ to suppress residual Galactic contamination. \textit{Middle:} Number overdensity of DESI Bright Galaxy Survey (BGS) sample, $\delta_\mathrm{g}$ ($\log \bar{M}_{200c}/M_\odot = 13.3$). \textit{Right:} ROSAT All-Sky Survey (RASS) X-ray-selected galaxy cluster sample~\citep{2023MNRAS.526.3757K}, colored and sized by cluster mass $M_{500c}$ ($\log \bar{M}_{200c}/M_\odot = 14.4$). These three catalogs form the tracer and rotator samples for the RM$^2$ stacking analysis.}
\label{fig:maps}
\end{figure*}

In this work, we show how RM catalogs, joint FRB+kSZ inferences on the cosmic baryon distribution, and tSZ measurements of thermal gas pressure can be combined with tracers of large-scale structure to derive constraints on (\textit{i}) the halo line-of-sight magnetic field $B_\parallel$, and (\textit{ii}) the plasma parameter $\beta$ in halos. First, we combine $\mathrm{RM}^2$ stacks of $\sim2\times10^5$ RMs from the second data release of the Spectral and Polarization In Cutouts of Extragalactic sources from RACS (SPICE-RACS DR2) survey~\citep{2026PASA...43...89T} on DESI Bright Galaxy Survey targets~\citep{2023AJ....165..253H} and ROSAT All-Sky Survey X-ray clusters~\citep{2023MNRAS.526.3757K}, with an independent measurement of the halo electron column densities from the FRB DM-redshift relation~\citep{2026NatAs.tmp..196S} and kSZ optical depth profiles~\citep{2026arXiv260419745H, 2026arXiv260419744Q}, and with the halo thermal pressure profile inferred from cosmic-shear cross-correlated with tSZ effect~\citep{2025arXiv250607432P}. We then present the first forecast for the DSA RM grid, and then forecast the improvement achievable with DSA-scale FRB and RM catalogs. This manuscript is organized as follows. We describe the datasets used in our analysis in Section~\ref{sec:data_description}. We present our cross-correlation measurements, their comparison with IllustrisTNG~\citep{2019ComAC...6....2N}, and our derived magnetic field and $\beta$ constraints in Section~\ref{sec:bfield}. We then present forecasts for DSA-scale datasets in Section~\ref{sec:forecast}, and discuss the implications and limitations of our results in Section~\ref{sec:discussion_and_conclusion}. Throughout this work, we use the \citet{2020A&A...641A...6P} cosmology, and define $M_\mathrm{200c}$ ($M_\mathrm{500c}$) as the total mass enclosed within a sphere of radius $R_\mathrm{200c}$ ($R_\mathrm{500c}$) whose mean interior density is 200 (500) times the critical density of the Universe.

\section{Data Description}
\label{sec:data_description}

In this section, we describe the data products used in our measurements, including the RM catalog used to construct the Faraday rotation field (Section~\ref{subsec:data_rm}), and the galaxy group and cluster tracer samples (Section~\ref{subsec:data_tracers}). The sky footprint of these three datasets is shown in Figure~\ref{fig:maps}. Additionally, to break the baryon-magnetism degeneracy, we use the electron column density constrained by FRB DM--redshift relation~\citep{2026NatAs.tmp..196S} and kSZ optical depth profiles~\citep{2026arXiv260419745H, 2026arXiv260419744Q} in our forthcoming study (Section~\ref{subsec:data_dm_tau}). To compare our measurements with thermal pressure in these halos, we use the profiles inferred from tSZ--cosmic shear cross-correlations (Section~\ref{subsec:data_pth}) by \citet{2025arXiv250607432P}.

\subsection{Rotation Measures from SPICE-RACS DR2}
\label{subsec:data_rm}

We use RMs from Data Release 2 of the SPICE-RACS survey~\citep{2026PASA...43...89T}, constructed from polarized source catalogs of the Australian Square Kilometer Array Pathfinder~\citep[ASKAP;][]{2021PASA...38....9H}. SPICE-RACS DR2 provides $246,509$ RMs measured at $\geq 8\sigma_p$ significance over 87.5\% of the sky south of declination $+49^\circ$, with a median RM uncertainty of 2.2~rad\,m$^{-2}$, and an on-sky source density of $\sim 6.7$~deg$^{-2}$, making it the widest-area RM catalog assembled to date. The images from which the RMs are extracted have an angular resolution of $\approx15\arcsec$. Because a source enters the catalog only if its polarized intensity exceeds the $\geq 8\sigma_p$ threshold, depolarization acts as a selection effect, removing sources behind strongly rotating or rapidly varying Faraday screens (Section~\ref{subsec:tng}). To isolate the extragalactic contribution from each measurement, we subtract a smoothed Galactic foreground RM map, constructed from the median RM of neighboring extragalactic sources on the sky,
\begin{equation}
    \mathrm{RRM}_i = \mathrm{RM}_i - \mathrm{RM}_\mathrm{MW}(\hat{n}_i),
    \label{eqn:rrm}
\end{equation}
where $\mathrm{RM}_\mathrm{MW}(\hat{n}_i)$ is evaluated by bilinear interpolation of the foreground map at each source position $\hat{n}_i$. We additionally discard sources at Galactic latitude $|b| < 25^\circ$ to suppress artifacts from residual Galactic contamination near the disk. After this cut, the maximum RRM in the sample is 564~rad\,m$^{-2}$. The resulting sky distribution of $|\mathrm{RRM}|$ is shown in the left panel of Figure~\ref{fig:maps}.

\subsection{DESI BGS Galaxy Groups and RASS Galaxy Clusters Samples}
\label{subsec:data_tracers}

We cross-correlate the residual RM field described above with two foreground tracer samples, selected to span roughly an order of magnitude in halo mass. 

Our lower-mass tracer is the Dark Energy Spectroscopic Instrument~\citep[DESI;][]{2019AJ....157..168D} Legacy Imaging Surveys Bright Galaxy Survey (BGS) bright sample, an $r<19.5$ magnitude-limited selection with color cuts to reject spurious detections, at photometric redshifts $0 < z < 0.5$~\citep{2023AJ....165..253H}. We use the Legacy Surveys mask to remove regions contaminated by bright stars, large galaxies, and globular clusters. This selection yields a sample of 15.4 million galaxies within the SPICE-RACS footprint with mean redshift $\bar{z} = 0.26$. The mean halo mass of this sample, computed using the halo occupation distribution calibrated to CMB lensing convergence around the same galaxies, is $\log \bar{M}_{200c}/M_\odot = 13.3$, tracing galaxy-group environments~\citep{2026arXiv260419745H}. 

Our higher-mass tracer is the ROSAT All-Sky Survey~\citep[RASS;][]{1993AdSpR..13l.391V} X-ray-selected galaxy cluster catalog~\citep{2023MNRAS.526.3757K}, constructed by applying a multi-component matched-filter cluster confirmation algorithm to the second RASS source catalog~\citep[2RXS;][]{2016A&A...588A.103B} and identifying optical counterparts via red-sequence galaxy overdensities. We restrict our analysis to clusters at Galactic latitude $|b| > 25^\circ$, yielding a sample of 7,775 clusters with mean redshift $\bar{z} = 0.18$ and mean halo mass $\log \bar{M}_{200c}/M_\odot = 14.4$, tracing galaxy-cluster environments, an order-of-magnitude more massive than the DESI BGS sample. 

\begin{figure*}[ht!]
\centering
\includegraphics[width=\textwidth]{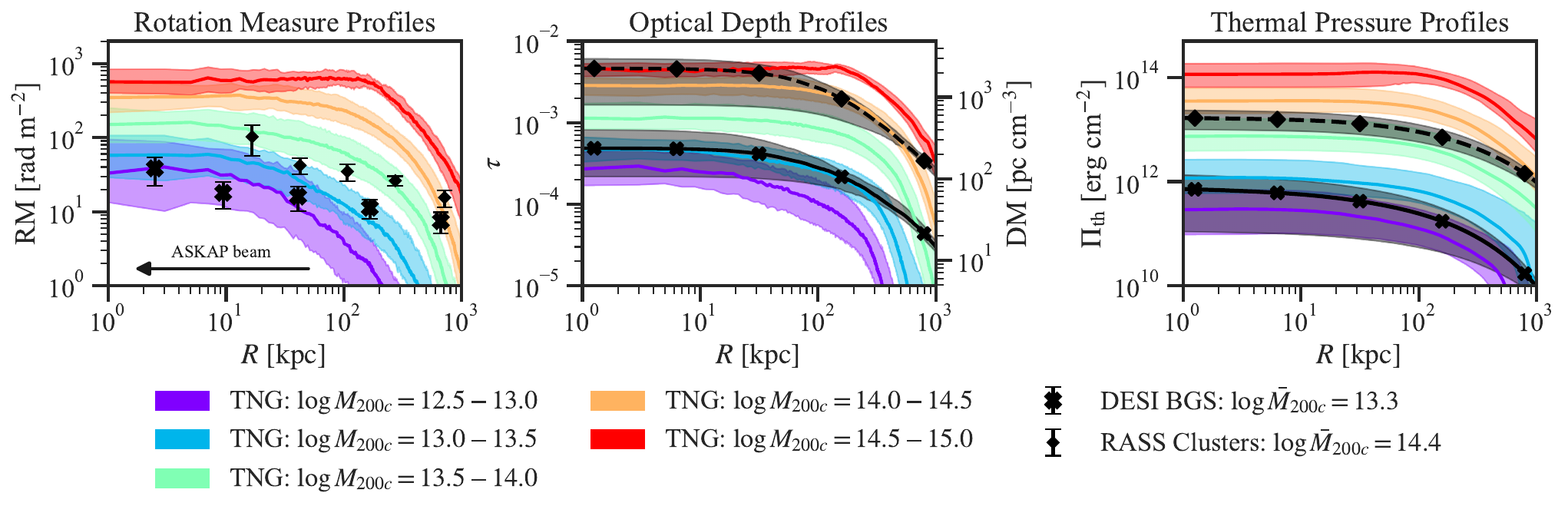}
\caption{Stacked profiles of RM (\textit{left}), optical depth / DM (\textit{middle}), and thermal pressure (\textit{right}) as a function of projected halo-centric radius $R$ around DESI BGS galaxies (crosses, $\log \bar{M}_{200c} = 13.3, \bar{z} = 0.26$) and RASS clusters (diamonds, $\log \bar{M}_{200c} = 14.4, \bar{z} = 0.18$). Colored curves and shaded bands show the median and 16th-84th percentile scatter of IllustrisTNG~\citep{2019ComAC...6....2N} profiles in five logarithmic mass bins spanning $\log (M_{200c}/M_\odot) = 12.5 - 15.0$. The left panel shows our measured RM$^2 \times$ tracer cross-correlations in rest-frame, with error bars showing $1\sigma$ jackknife uncertainties. The arrow marks the scales affected by the ASKAP beam resolution ($R\lesssim50$~kpc, set by the $\approx15\arcsec$ beam), which any modeling of these measurements must account for. The middle panel shows our $\tau$/DM profiles computed using posterior samples of the gas distribution parameters constrained by FRB DM--redshift relation~\citep{2026NatAs.tmp..196S} and ACT-DR6 $\times$ DESI-DR2 kSZ optical depth profile measurements~\citep{2026arXiv260419744Q, 2026arXiv260419745H}. The right panel shows the projected thermal pressure profiles inferred from ACT-DR6 $\times$ DES-Y3 tSZ--cosmic shear cross-correlations~\citep{2025arXiv250607432P}.}
\label{fig:RM_tau_Pt}
\end{figure*}

The sky distribution of the DESI BGS number overdensity and the RASS cluster sample are shown in the middle and right panels of Figure~\ref{fig:maps}, respectively.

\subsection{Electron Column Density from FRB DM--$z$ Relation and kSZ Optical Depth Profiles}
\label{subsec:data_dm_tau}

To break the baryon-magnetism degeneracy of Equation~\ref{eqn:RM}, we require an electron density estimate for the DESI BGS and RASS halo populations that is independent of the RM measurement. We compute these halo gas profiles using the analytical \citet{2019JCAP...03..020S} model implemented in \sw{BaryonForge}\footnote{\url{https://github.com/DhayaaAnbajagane/BaryonForge}}~\citep{2024OJAp....7E.108A}, with the profile parameters drawn from the posterior of the self-consistent analysis of Sharma et al. (in prep), which jointly fits for the FRB DM--redshift relation~\citep{2026NatAs.tmp..196S} and the Atacama Cosmology Telescope (ACT) DR6 $\times$ DESI DR2 kSZ optical depth profiles. We propagate the posterior scatter on feedback parameters, computing, for 1,000 posterior samples, the projected gas surface density $\Sigma_\mathrm{gas}(R)$ of a $10^{13.3}M_\odot$ halo at $z = 0.26$ (BGS) and $10^{14.4}M_\odot$ halo at $z = 0.18$ (RASS), and converting it to an optical depth profile $\tau(R)=\sigma_\mathrm{T}\chi_\mathrm{e}\Sigma_\mathrm{gas}(R)/m_\mathrm{p}$ and electron column density $N_\mathrm{e}(R)=\tau(R)/\sigma_\mathrm{T}$, where $\sigma_\mathrm{T}$ is the Thomson scattering cross-section, $m_\mathrm{p}$ is the proton mass and $\chi_\mathrm{e}=1-Y_\mathrm{p}/2=0.877$ for a primordial helium mass fraction $Y_\mathrm{p}=0.2453$. This gives the median profile and 16th--84th percentile uncertainty band, shown in the middle panel of Figure~\ref{fig:RM_tau_Pt}.

\subsection{Thermal Pressure from tSZ--Cosmic Shear Cross-Correlations}
\label{subsec:data_pth}

The halo thermal pressure profiles are obtained from the joint analysis of Dark Energy Survey (DES) Y3 cosmic shear and ACT DR6+Planck tSZ Compton-$y$ observations by \cite{2025arXiv250607432P}. That work detects their cross-correlation at 21$\sigma$, and jointly models it with the small-scale shear auto-correlation in a halo-model framework with a parameterized gas pressure profile, constraining the thermal pressure of halos as a function of mass and redshift. We obtained the posterior samples of that analysis from S. Pandey via private communication. For each posterior sample, we computed the total thermal gas pressure profile, $P_\mathrm{th}(r)$ (electrons plus ions, $P_\mathrm{th} = 1.932 P_\mathrm{e}$), of halos with the characteristic masses and redshifts of our two tracer samples, and we use the posterior median and the 68\% confidence interval. Since RM and $\tau$ are both line-of-sight integrals, we project $P_\mathrm{th}$ along the line-of-sight $\Pi_\mathrm{th}(R)=\int P_\mathrm{th}\left(\sqrt{R^2+l^2}\right)\mathrm{d}l$, in physical units. These projected profiles are shown in the right panel of Figure~\ref{fig:RM_tau_Pt}.

Combining this thermal pressure with the electron density of Section~\ref{subsec:data_dm_tau}, the implied gas temperature, $k_\mathrm{B}T(r)=P_\mathrm{th}(r)/[1.932\,n_\mathrm{e}(r)]$, is lower than the virial expectation for these halos by a factor of $\approx2.5$--$4$ for the RASS clusters and by a factor of $\approx3$--$5$ for the BGS groups. This is consistent with the tSZ deficit, and with the degeneracy between strong feedback and non-thermal pressure support that affects tSZ inferences (Section~\ref{sec:introduction}).

\section{RM\,$^2$ Stacks and Magnetic Field Constraints}
\label{sec:bfield}

In this section, we measure the RM$^2\times$tracer cross-correlations (Section~\ref{subsec:bfield_estimator}), compare the resulting profiles with IllustrisTNG halos (Section~\ref{subsec:tng}), and combine them with the electron column density and thermal pressure profiles of Section~\ref{sec:data_description} to derive limits on the line-of-sight magnetic field strength (Section~\ref{subsec:b_estimator}) and the plasma $\beta$ parameter (Section~\ref{subsec:plasma_beta}).

\subsection{RM\,$^2\times$Tracer Cross-Correlation Measurement}
\label{subsec:bfield_estimator}

\begin{figure*}[ht!]
\centering
\includegraphics[width=\textwidth]{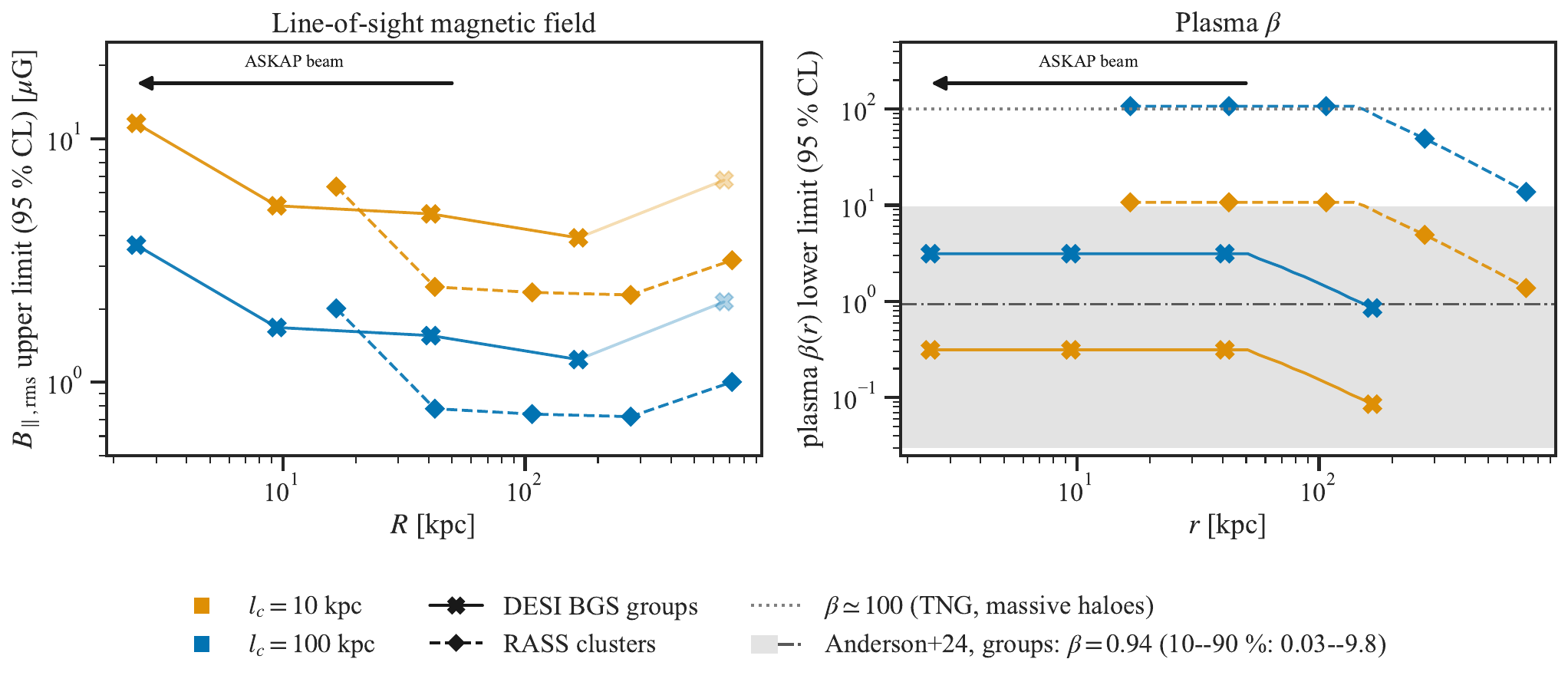}
\caption{Current constraints on the magnetic field strength in DESI BGS galaxy groups and RASS galaxy clusters obtained from the stacked RM$^2$ combined with electron density~\citep{2026NatAs.tmp..196S, 2026arXiv260419745H, 2026arXiv260419744Q} and thermal pressure~\citep{2025arXiv250607432P} constraints from literature. \textit{Left}: 95\% upper limits on the rms line-of-sight field for two assumed coherence lengths of 10 and 100~kpc (limits scale as $l_c^{-1/2}$); bins beyond $R_{200c}$ are faint. \textit{Right}: 95\% lower limits on $\beta(r)$, marginalized over the uncertainties in the RM$^2$ stacks, and electron density and thermal pressure literature constraints, showing at each radius, the weaker of a constant-$\beta$ and a power-law-$\beta(r)$ model (limits scale as $l_c$). The gray band and dash-dotted line show the median and 10th--90th percentile plasma $\beta$ of the intragroup medium of 55 nearby groups estimated from POSSUM RMs by \citet{2024MNRAS.533.4068A}, an indicative estimate averaged within two splashback radii. In both panels, the left-facing arrow marks the scales, $R\lesssim50$~kpc, affected by the ASKAP beam resolution (the $\approx15\arcsec$ beam is $\approx47$--$63$~kpc at the sample redshifts). The coherence length is not measured, and field reversals and selection effects of polarized sources, notably depolarization, can bias the RM$^2$ low. The weaker upper-limits in clusters than in groups is likely an indication that cluster sightlines are affected by such effects rather than evidence for weak cluster fields.}
\label{fig:B_beta_limits}
\end{figure*}

We stack the squared residual rotation measure, $\mathrm{RRM}^2$ (Equation~\ref{eqn:rrm}), around the BGS and RASS tracer samples. Compared to stacking RM or $|\mathrm{RM}|$, stacking RM$^2$ is not affected by signal cancellation from field reversals along the line-of-sight or by the noise bias of $|\mathrm{RM}|$, and has a direct interpretation in terms of the electron-density-weighted line-of-sight magnetic field~\citep{2026JCAP...07..096Z}. We define the stack as the excess mean RM$^2$ of polarized sources at projected physical radius $R$ from a tracer,
\begin{equation}
    \boldsymbol{\xi}(R) = \langle \mathrm{RRM}^2 \rangle_{R} - \langle \mathrm{RRM}^2 \rangle_{\mathrm{randoms}},
    \label{eqn:xi_RM2}
\end{equation}
where $\langle \mathrm{RRM}^2 \rangle_{R}$ is the mean $\mathrm{RRM}^2$ of the sources in an annulus at projected radius $R$ around the tracers, $\langle \mathrm{RRM}^2 \rangle_{\mathrm{randoms}}$ is the same quantity around random positions in the survey footprint, and $\boldsymbol{\xi}$ is in units of rad$^2$\,m$^{-4}$ in the observer-frame. We measure $\boldsymbol{\xi}$ with \sw{TreeCorr}~\citep{2015ascl.soft08007J}, following the cross-correlation methodology of \citet{2026ApJ..1010...47S, 2026ApJ..1006L...3S}, in $N_\mathrm{bin}=5$ logarithmically spaced bins of projected physical separation spanning 1--1,000~kpc for BGS and 10--1,000~kpc for RASS samples. The cross-correlation down to these small scales is only limited by the number of source--tracer pairs, which we ensure is $\geq500$ in every bin to provide sufficient statistics. The ASKAP beam does, however, affect the measured profile on small scales ($R\lesssim50$~kpc; arrows in Figures~\ref{fig:RM_tau_Pt} and~\ref{fig:B_beta_limits}). This scale is set by the SPICE-RACS DR2 point-spread-function (PSF), whose $\approx15\arcsec$ full width at half maximum corresponds to $\approx63$~kpc at $\bar{z}=0.26$ (BGS) and $\approx47$~kpc at $\bar{z}=0.18$ (RASS), or $\approx0.12R_{200c}$ and $\approx0.04R_{200c}$. At smaller separations, the beam can bias the stack in two opposing ways. Smoothing of RM structure within the beam, and depolarization by RM gradients across it that removes sightlines with the largest RM (Section~\ref{subsec:tng}), bias the stack low, so that these bins trace a beam-convolved profile. Conversely, the polarized source and the tracer can be the same object or unresolved blends, so that the RM reflects Faraday rotation local to the radio source rather than halo gas, which biases the stack high. The net effect is not known, so the bins at $R\lesssim50$~kpc cannot be compared directly with a model of the halo gas. The tightest field bounds in each sample, $3.9~\mu$G for BGS ($R=166$~kpc) and $2.3~\mu$G for RASS ($R=273$~kpc), come from bins outside this region (Section~\ref{subsec:b_estimator}). Any modeling that uses these measurements must therefore account for the beam shape, for example by convolving the model with the beam.

We estimate the covariance, $\mathbf{C}$, by jackknife resampling over $N_\mathrm{patch}=100$ spatial patches. The covariance is further averaged over 100 random jackknife patch realizations. The total signal-to-noise ratio is $\mathrm{S/N}=\sqrt{f_\mathrm{H}\boldsymbol{\xi}^\mathrm{T}\mathbf{C}^{-1}\boldsymbol{\xi}-N_\mathrm{bin}}$, where $f_\mathrm{H}=(N_\mathrm{patch}-N_\mathrm{bin}-2)/(N_\mathrm{patch}-1)$ is the \citet{2007A&A...464..399H} correction for noise in the inverse of the covariance matrix. We obtain $\mathrm{S/N}=1.9$ ($3.6$) for the BGS (RASS) stacks.

To compare with magneto-hydrodynamical simulations, we convert to the rest-frame RM amplitude, $\mathrm{RM}(R)=(1+\bar{z})^2\sqrt{\boldsymbol{\xi}(R)}$, where $\bar{z}$ is the mean redshift of the sample. The resulting profiles are shown in the left panel of Figure~\ref{fig:RM_tau_Pt}, with $1\sigma$ jackknife uncertainties. The null tests, in which the RMs are randomly shuffled among the sources, are presented in Appendix~\ref{app:null_tests}. The profiles at larger radii, where the signal declines toward zero, are shown in Appendix~\ref{app:large_radii}.

\subsection{Comparison with IllustrisTNG}
\label{subsec:tng}

To place the measured profiles in context, we compare them to halos in the TNG300 magneto-hydrodynamical simulation of the IllustrisTNG project~\citep{2018MNRAS.480.5113M, 2019ComAC...6....2N}. We use the $z=0.3$ snapshot and halos in five $0.5$~dex bins of $\log(M_{200c}/M_\odot)$ from $12.5$ to $15.0$. For each halo, we project the gas electron density and magnetic field along three orthogonal axes onto $2$~kpc pixel maps, compute the corresponding RM and optical depth maps, and measure the azimuthal mean of $|\mathrm{RM}|$ and of $\tau$ in annuli, as well as the thermal pressure of the non-star-forming gas projected along the line of sight. Figure~\ref{fig:RM_tau_Pt} shows the median and the 16th--84th percentile scatter among halos in each mass bin.

The $\tau$ profiles of both samples, which set the electron column density, are broadly consistent with those of TNG halos of similar mass. The BGS RM profile is likewise comparable in amplitude to TNG halos of group mass at $R<200$~kpc. In contrast, the RASS RM profile lies well below TNG halos of cluster mass at all radii, despite the agreement in electron column density. The TNG profiles show the mean $|\mathrm{RM}|$, whereas the data give the rms, $\sqrt{\boldsymbol{\xi}}$, but for a Gaussian distribution of RM these differ only by a factor $\sqrt{2/\pi}\simeq0.8$, much smaller than the RASS offset. 

The RM amplitude of the RASS stack may be biased low because Faraday depolarization acts as a selection effect, which removes sources behind strongly rotating and rapidly varying Faraday screens from the polarized source catalog. Other selection effects arise because a source enters the polarized catalog only if its polarized intensity exceeds the $8\sigma_p$ detection threshold and its Faraday spectrum yields a reliable RM. Sources whose polarization is reduced by RM gradients across the beam or across a spectral channel can drop out, and sightlines with large $|\mathrm{RM}|$ or complex, multi-component Faraday spectra are harder to measure and may be removed by quality cuts. If the detection probability decreases with the Faraday depth or dispersion of the screen, the surviving sources preferentially trace low-RM sightlines, biasing the RM$^2$ stack low where the gas is densest and most magnetized. Unknown source redshifts further dilute the stacked signal due to sources in the foreground or sources embedded in the tracer samples. Quantifying these effects is beyond the scope of this work. We therefore interpret the field strengths and plasma $\beta$ derived below as bounds on the field probed by the surviving polarized sources, conditional on $l_c$, rather than as measurements of the physical field.

\subsection{From RM\,$^2$ and $\tau$ to $B_\parallel$} \label{subsec:b_estimator}

We combine the stack with the electron column density to estimate the line-of-sight magnetic field. In each radial bin, we compute the rest-frame DM from the optical depth profile of Section~\ref{subsec:data_dm_tau}, $\mathrm{DM}=\tau/(\sigma_\mathrm{T}\,\mathrm{pc})$ in pc\,cm$^{-3}$, and the field of a coherent uniform screen that would produce the measured RM,
\begin{equation}
    B_{\parallel,\mathrm{eff}}(R) = \frac{\mathrm{RM}_\mathrm{rest}(R)}{0.812\,\mathrm{DM}(R)}\ \mu\mathrm{G},
    \label{eqn:Beff}
\end{equation}
where $\mathrm{RM}_\mathrm{rest}$ is in rad\,m$^{-2}$ and $0.812$ is the RM--DM constant in rad\,m$^{-2}$ per $\mu$G\,pc\,cm$^{-3}$. Because the field along a sightline reverses on a coherence length $l_c$, the actual field is stronger. For a sightline of length $L$ crossing $L/l_c$ independent cells with randomly oriented fields, the RM contributions of the cells add in quadrature while their DM contributions add linearly, so that the rms line-of-sight field is
\begin{equation}
    B_{\parallel,\mathrm{rms}}(R) = B_{\parallel,\mathrm{eff}}(R)\sqrt{\frac{L(R)}{l_c}},
    \label{eqn:Brms}
\end{equation}
where $L(R)=2\sqrt{(2R_{200c})^2-R^2}$ is the chord length through a sphere of radius $2R_{200c}$, with $R_{200c}=0.53$ and $1.26$~Mpc for BGS and RASS, respectively. Equation~\ref{eqn:Brms} assumes a uniform electron density along the sightline, a simplification that is relaxed in Section~\ref{subsec:plasma_beta}.

The stacks are detected at total significances of $1.9\sigma$ (BGS) and $3.6\sigma$ (RASS), but only at modest significance in individual radial bins, and the conversion to a field strength depends on the unmeasured coherence length. We therefore report one-sided 95\% upper bounds on $B_{\parallel,\mathrm{rms}}$ instead of point estimates, i.e., the 95th percentile of $10^4$ joint Monte Carlo draws of $\boldsymbol{\xi}$ in each bin and of the $\tau$ posterior samples of Section~\ref{subsec:data_dm_tau}. For $l_c=10$~kpc (Figure~\ref{fig:B_beta_limits}, left), we find $B_{\parallel,\mathrm{rms}}<3.9$--$11.6~\mu$G for the BGS groups and $<2.3$--$6.4~\mu$G for the RASS clusters at $R<R_{200c}$. The limits scale as $l_c^{-1/2}$, and are a factor $\sqrt{10}$ tighter for $l_c=100$~kpc (Figure~\ref{fig:B_beta_limits},~left). At fixed $l_c$, the RASS limits are a factor of $\sim2$ tighter than the BGS limits, because the RM amplitude rises less steeply from groups to clusters than the electron column density does. Together with the comparison to IllustrisTNG in Section~\ref{subsec:tng}, this suggests that the RASS stack is likely affected by depolarization rather than that the cluster fields are weak. Therefore, the limits refer to the RM$^2$ amplitude that survives in the polarized-source stack: if depolarization removes the sightlines with the largest RM, the physical field can be stronger than these limits.

We note that the coherence length is not constrained by our data, but the literature offers guidance for clusters. RM maps of radio sources embedded in cluster cores show patches with coherence lengths of 5--10~kpc in cooling-flow clusters and 15--30~kpc in non-cooling-flow clusters~\citep{2002ARA&A..40..319C}, and the field in the Coma cluster is described by a Kolmogorov spectrum spanning $\sim2$--$34$~kpc~\citep{2010A&A...513A..30B}. Ordered components on $\sim100$~kpc scales have also been seen, for example in Hydra~A~\citep{2002ARA&A..40..319C}. Overall, the field structure in clusters remains poorly known~\citep{2019SSRv..215...16V}, and we are not aware of direct measurements in galaxy groups. A few to a few tens of kpc therefore appears more representative than $l_c=100$~kpc, and thus we quote the $l_c=10$~kpc limits, as the more conservative case (Figure~\ref{fig:B_beta_limits}, left).

Additionally, these limits are conditional on the assumptions of an isotropic random field and a sightline truncated at $2R_{200c}$. Bins at $R>R_{200c}$ are shown faint in Figure~\ref{fig:B_beta_limits}, as the halo gas model is not expected to well-describe the column density there (two-halo contribution is expected to be significant at large scales). The RM profile at such large radii is shown in Appendix~\ref{app:large_radii}.

\subsection{Magnetic Pressure and Plasma $\beta$}
\label{subsec:plasma_beta}

The plasma $\beta = P_\mathrm{th}/P_B$ compares the thermal pressure, $P_\mathrm{th}$, of Section~\ref{subsec:data_pth} to the magnetic pressure, $P_B=B^2/8\pi$. For an isotropic random field, $B_{\parallel,\mathrm{rms}}^2=B^2/3$, and therefore $B_{\parallel,\mathrm{rms}}^2=(8\pi/3)\,P_\mathrm{th}/\beta$. Since $\beta$ is defined at three-dimensional radius $r$, whereas the stack is a projected quantity, we forward-model the stack from the three-dimensional profiles instead of inverting it. In the cell model of Section~\ref{subsec:b_estimator}, the RM$^2$ stack is
\begin{equation}
    \boldsymbol{\xi}(R) = \frac{0.812^2\,(8\pi/3)\,10^{12}}{(1+\bar{z})^4}\; l_c \int n_\mathrm{e}^2(r)\,\frac{P_\mathrm{th}(r)}{\beta(r)}\,\mathrm{d}l,
    \label{eqn:xi_model}
\end{equation}
where $r=\sqrt{R^2+l^2}$, $n_\mathrm{e}(r)$ is the free electron density in cm$^{-3}$, $P_\mathrm{th}(r)$ is in erg\,cm$^{-3}$, $l$ and $l_c$ are in pc, the factor $10^{12}$ converts G$^2$ to $\mu$G$^2$, the factor $(1+\bar{z})^{-4}$ converts to the observed frame, and the integral along the line of sight runs over $|l|\leq\sqrt{(2R_{200c})^2-R^2}$. We use the three-dimensional $n_\mathrm{e}(r)$ from the same posterior samples as $\tau$ (Section~\ref{subsec:data_dm_tau}).

We consider two models for the radial dependence of $\beta$: a constant, $\beta(r)=\beta_p$, and a power law, $\beta(r)=\beta_p\,(r/r_p)^{\alpha}$, with $r_p=0.1R_{200c}$ and a flat prior $-2\leq\alpha\leq2$. In both models, the predicted stack is linear in $x\equiv l_c/\beta_p$, so the data constrain only the combination $l_c/\beta_p$. With a flat prior on $x\geq0$ and a Gaussian likelihood with the jackknife covariance (corrected by $f_\mathrm{H}$), the posterior of $x$ is a truncated Gaussian, and the 95\% lower limit on $\beta_p$ is $l_c/x_{95}$, with $x_{95}$ the 95th percentile of $x$. We marginalize over the posterior samples of $n_\mathrm{e}$, over $P_\mathrm{th}$, for the power-law model, over $\alpha$. We fit only the bins at $R<R_{200c}$: four for BGS and five for RASS.

For a constant $\beta$, the 95\% lower limits are $\beta>0.31$ for BGS, and $\beta>11$ for RASS, assuming $l_c=10$~kpc. Allowing a power-law radial dependence weakens the limits at radii away from the pivot radius, and Figure~\ref{fig:B_beta_limits} (right panel) shows the weaker of the two limits at each radius. For $l_c=10$~kpc, this gives $\beta(r)>0.09$--$0.3$ for BGS and $\beta(r)>1.4$--$11$ for RASS over the fitted radii, and all limits scale linearly with $l_c$. For comparison, the dotted line in Figure~\ref{fig:B_beta_limits} marks $\beta\simeq100$, as found in massive IllustrisTNG halos~\citep[$M_{200c} \gtrsim 10^{13.5}M_\odot$;][]{2018MNRAS.480.5113M}. For galaxy groups, \citet{2024MNRAS.533.4068A} infer a median $\beta\approx0.9$, with a 10th--90th percentile range of $0.03$--$9.8$ (gray band), from RM dispersion measurements around 55 nearby groups ($z\leq0.025$, $10^{12.5}$--$10^{14}M_\odot$). These estimates are indicative, relying on assumed electron densities, temperatures, and coherence scales. Their median is consistent with our BGS limit, which still permits magnetically dominated groups. They also find groups to be more magnetically dominated than the intracluster medium, with typical $\beta\sim10$--$100$, consistent with our RASS limit. More generally, $\beta=100$ remains consistent with the RASS stack only for $l_c\lesssim90$~kpc, while the BGS stack excludes this value only for $l_c\gtrsim3$~Mpc. As for $B_\parallel$, these constraints are conditional on $l_c$ and subject to the depolarization selection effect discussed in Section~\ref{subsec:tng}.

\section{Opportunities with DSA-Scale Faraday Rotation and DM Datasets} \label{sec:forecast}

The constraints of Section~\ref{subsec:plasma_beta} are limited by the S/N of the RM$^2$ stacks, kSZ effect stacks, and sample size of FRBs with redshift measurements. These limitations will be addressed by upcoming surveys: orders of magnitude larger catalogs of RMs from the DSA~\citep{2019BAAS...51g.255H} will reduce the statistical uncertainty of the RM$^2$ stacks, while localized FRBs from DSA will measure the electron column density directly through the FRB DM$\times$galaxy cross-correlation. In this section, we forecast the resulting precision on the line-of-sight magnetic field of the BGS galaxy groups.

\subsection{DSA RM grid forecast}
\label{subsec:dsa_rm}

The DSA comprises 1650 antennas of 6.15\,m diameter, observing from 0.70--2.00\,GHz in two linear
polarizations. Its Cadenced All-Sky Survey (CASS) covers the 30,940\,deg$^2$ north of $\delta = -30^\circ$ in multiple epochs of 21-minute pointings. Stokes Q/U image cubes are produced with 500 channels of 2.604\,MHz width. For an example 1-hr pointing on a source transiting near zenith, the band-averaged noise in Stokes $Q$ or $U$, which is also the noise of the Faraday-depth spectrum from RM synthesis with inverse-variance channel weights, is $\sigma_{Q,U} \approx 0.64\,\mu$Jy\,beam$^{-1}$ with natural weighting ($12\,\mu$Jy\,beam$^{-1}$ per 2.6-MHz channel at 1.4\,GHz).\footnote{We use the DSA sensitivity calculator, \url{https://www.deepsynoptic.org/sensitivity/index.html}, for all DSA sensitivities in this section.}

One possible strategy for an RM grid matched to the BGS is to stack seven CASS epochs in the northern Euclid \citep{2022A&A...662A.112E} and HSC \citep{2018PASJ...70S...4A} fields (10,606\,deg$^2$ in total in CASS). The mosaicked cubes reach a median $\sigma_p = 0.36\,\mu$Jy\,beam$^{-1}$ (10th--90th percentile 0.34--0.40\,$\mu$Jy\,beam$^{-1}$) with robust ($R = 0.4$) imaging weighting, or $0.32\,\mu$Jy\,beam$^{-1}$ with natural weighting. Channels that are uniform in frequency crowd toward short $\lambda^2$, whereas the mosaic noise is lowest at long $\lambda^2$, where the primary beam is largest. With inverse-variance channel weights the effective RMSF FWHM is 25.5\,rad\,m$^{-2}$ (33\,rad\,m$^{-2}$ for equal channel weights). 

We estimate the RM source density from the 1.4~GHz polarized source counts of \citet{2019MNRAS.485.5285L}, $N(>p) = 2.01\,(p/{\rm mJy})^{-0.89}$\,deg$^{-2}$, requiring $\geq 8\sigma_p$ in the Faraday-depth spectrum as for SPICE-RACS (Section~\ref{subsec:data_rm}). The band-averaged Faraday spectrum refers to $\nu \approx 1.0$\,GHz, where a source with spectral index $\alpha = -0.7$ appears 21\% brighter than at 1.4\,GHz. However, because Faraday and beam depolarization reduce the polarized fraction of many sources toward low frequencies, we conservatively adopt a flat polarized spectrum ($\alpha = 0$). This gives 365 RMs\,deg$^{-2}$, with a range of 190--690\,deg$^{-2}$ from the uncertainties on the normalization and slope of the counts, compared with $\sim 6.7$\,deg$^{-2}$ for SPICE-RACS. At $5\sigma_p$ with $\alpha = -0.7$ the density rises to 660\,deg$^{-2}$, which may be directly compared with the results of \citet{2026aask.bookE.132L} for fiducial SKAmid targeted and mosaicked observations. The Euclid and HSC fields overlap the BGS sample over 9,961\,deg$^2$ ($f_{\rm sky} \approx 0.24$), which contains $\approx 9.7 \times 10^6$ BGS galaxies at the mean density of our sample and $\approx 3.6 \times 10^6$ RMs (range $1.9$--$6.9 \times 10^6$).

\subsection{Forecast Formalism}
\label{sec:forecast_formalism}

Since $B_{\parallel,\mathrm{eff}}\propto\sqrt{\boldsymbol{\xi}}/\mathrm{DM}$ (Equation~\ref{eqn:Beff}), the fractional uncertainty on the line-of-sight field in each radial bin is
\begin{equation}
    \left(\frac{\sigma_B}{B}\right)^2 = \left(\frac{\sigma_\xi}{2\,\boldsymbol{\xi}}\right)^2 + \left(\frac{\sigma_\mathrm{DM}}{\mathrm{DM}}\right)^2,
    \label{eqn:sigma_lnB}
\end{equation}
\begin{figure*}[ht!]
\centering
\includegraphics[width=\textwidth]{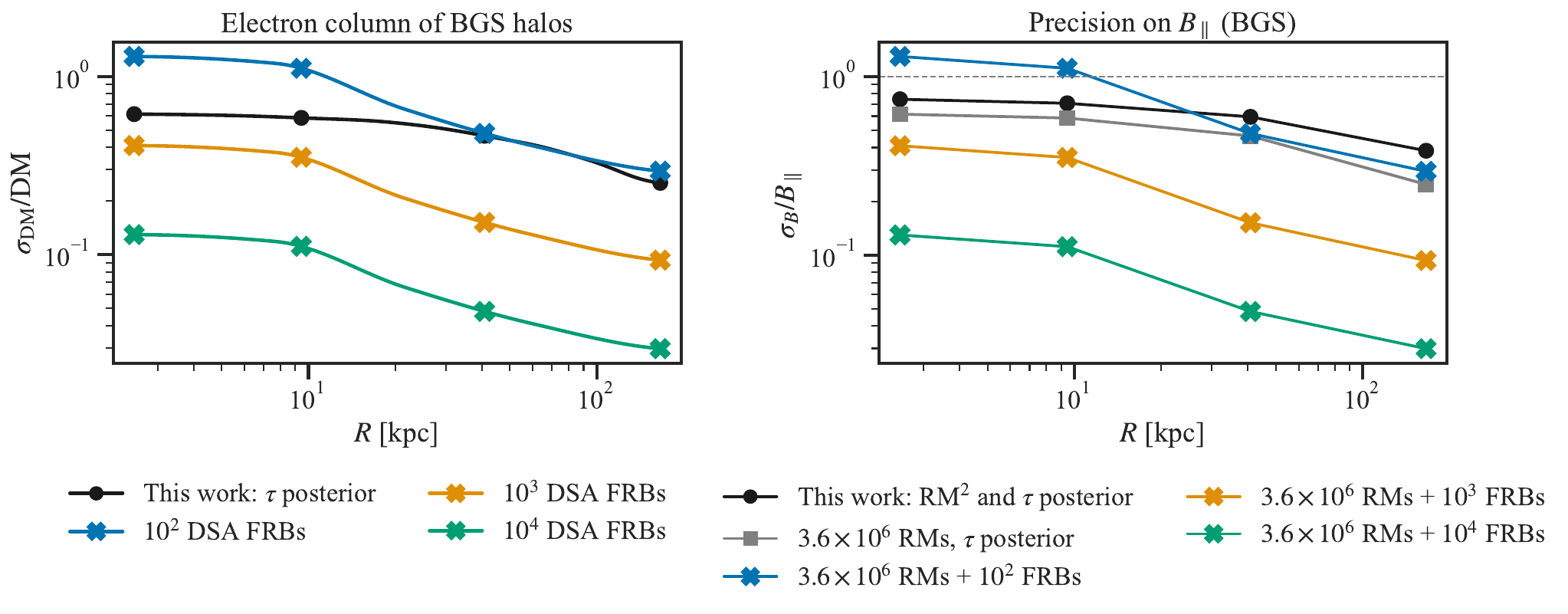}
\caption{Forecasted improvement from DSA FRBs and RMs for DESI BGS halos. \textit{Left}: Current fractional precision on the DM profile out to the outermost RM-stack bin inside $R_{200c}$~\citep{2026NatAs.tmp..196S, 2026arXiv260419745H, 2026arXiv260419744Q}, and forecast for DSA FRB $\times$ galaxy cross-correlations with $10^2$, $10^3$ and $10^4$ FRBs (colored curves). \textit{Right}: Current fractional precision on the line-of-sight magnetic field $B_\parallel=\mathrm{RM}/(0.812\,\mathrm{DM})$, alongside a forecast for $3.6 \times 10^6$ DSA RMs combined with the DSA FRB samples.}
\label{fig:DM_RM_forecast}
\end{figure*}
where $\sigma_\xi$ is the uncertainty on the RM$^2$ stack and $\sigma_\mathrm{DM}/\mathrm{DM}$ is the fractional uncertainty on the DM profile. We hold $\boldsymbol{\xi}$ at its measured value, so that the forecast quantifies only the reduction of the statistical errors. The result is the precision on $B_{\parallel,\mathrm{eff}}$, the combination that is independent of the coherence length $l_c$. For the present-day precision, $\sigma_\xi$ is the jackknife uncertainty of Section~\ref{subsec:bfield_estimator}, and $\sigma_\mathrm{DM}/\mathrm{DM}$ is the scatter of the $\tau$ posterior of Section~\ref{subsec:data_dm_tau}. For the forecasts, we replace them with the expected uncertainties for $N_\mathrm{RM}$ RMs and $N_\mathrm{FRB}$ localized FRBs, respectively.

We forecast the FRB DM$\times$galaxy cross-correlation with the framework of \citet{2026ApJ...998..109S}, to which we refer the reader for details. For a galaxy field $g$ and a field $X$, the Gaussian covariance of the angular cross-power spectrum~\citep{2017MNRAS.470.2100K} is
\begin{equation}
    \mathrm{Var}\left[C_\ell^{gX}\right] = \frac{\left(C_\ell^{gX}\right)^2 + \left(C_\ell^{gg}+N_g\right)\left(C_\ell^{XX}+N_X\right)}{(2\ell+1)\,f_\mathrm{sky}},
    \label{eqn:var_Cl}
\end{equation}
where $\ell$ is the multipole, $C_\ell^{gX}$, $C_\ell^{gg}$ and $C_\ell^{XX}$ are the cross- and auto-power spectra, $N_g$ and $N_X$ are the shot-noise power spectra of the galaxies and of $X$, and $f_\mathrm{sky}$ is the sky fraction. For the FRB forecast, $X=\mathrm{DM}$, and the spectra are the halo-model spectra for the BGS sample. For the RM forecast, we follow \citet{2026JCAP...07..096Z} and assume that the RM$^2$ field is shot-noise dominated, such that $\mathrm{Var}[C_\ell^{g\mathrm{RM}^2}]=(C_\ell^{gg}+N_g)\,N_{\mathrm{RM}^2}/[(2\ell+1)f_\mathrm{sky}]$. The shot-noise power spectra are
\begin{equation}
    \begin{split}
    N_g &= \frac{\Omega}{N_\mathrm{gal}}, \\
    N_\mathrm{DM} &= \frac{\sigma_\mathrm{host}^2}{(1+\bar{z}_\mathrm{FRB})^2}\,\frac{\Omega}{N_\mathrm{FRB}}, \\
    N_{\mathrm{RM}^2} &= \sigma_\mathrm{RM}^4\,\frac{\Omega}{N_\mathrm{RM}},
    \end{split}
    \label{eqn:shot_noise}
\end{equation}
where $\Omega=4\pi f_\mathrm{sky}$ is the survey solid angle, $N_\mathrm{gal}$, $N_\mathrm{FRB}$ and $N_\mathrm{RM}$ are the number of galaxies, localized FRBs and RMs, $\sigma_\mathrm{host}$ is the scatter of the host-galaxy DM contribution, $\bar{z}_\mathrm{FRB}$ is the mean FRB redshift, and $\sigma_\mathrm{RM}$ is the effective RM scatter per source.

We adopt $f_\mathrm{sky}=0.24$, the overlap of the DSA RM grid with the DESI BGS sample (Section~\ref{subsec:dsa_rm}), for which $\Omega=3.02$~sr, and assume that the galaxies, FRBs and RMs cover the same footprint. For the DM forecast, we use $N_\mathrm{gal}=9.7\times10^6$ galaxies ($N_g=3.1\times10^{-7}$~sr), $\sigma_\mathrm{host}=100$~pc\,cm$^{-3}$, and an FRB redshift distribution $\mathrm{d}N/\mathrm{d}z\propto z^2e^{-3.5z}$ with $\bar{z}_\mathrm{FRB}=0.86$~\citep{2026ApJ...998..109S}. For $N_\mathrm{FRB}=10^2$, $10^3$ and $10^4$, this gives $N_\mathrm{DM}=87.5$, $8.8$ and $0.9$~(pc\,cm$^{-3}$)$^2$\,sr, respectively. For the RM forecast, we use $\sigma_\mathrm{RM}=14.3$~rad\,m$^{-2}$, calibrated to reproduce the jackknife uncertainties of the present stack, which gives $N_{\mathrm{RM}^2}=0.035$~(rad\,m$^{-2}$)$^4$\,sr for $N_\mathrm{RM}=3.6 \times 10^6$ (the DSA RM grid forecast of Section~\ref{subsec:dsa_rm}).

\subsection{Forecast Results}
\label{sec:forecast_results}

Figure~\ref{fig:DM_RM_forecast} (left) shows the fractional precision on the electron column density profile. The present constraint from the posterior of Section~\ref{subsec:data_dm_tau} corresponds to $\sigma_\mathrm{DM}/\mathrm{DM}\simeq0.25$--$0.6$. The FRB DM$\times$galaxy cross-correlation measures the electron column density directly, and already for $10^2$ localized FRBs, the precision is comparable to the present constraints at $R\geq40$~kpc, with $\sigma_\mathrm{DM}/\mathrm{DM}=0.48$, $0.34$ and $0.29$ at $R=40$, $100$ and $170$~kpc, compared to $0.47$, $0.34$ and $0.25$ today. The precision then improves as $\propto N_\mathrm{FRB}^{-1/2}$, reaching $0.15$, $0.11$ and $0.09$ for $10^3$ FRBs and $0.05$, $0.03$ and $0.02$ for $10^4$ FRBs at the same radii.

Figure~\ref{fig:DM_RM_forecast} (right) shows the resulting precision on $B_\parallel$. In our work, the stack has a per-bin RM$^2$ S/N of only $1.2$--$1.7$, and $\sigma_B/B=0.38$--$0.75$. With $3.6 \times 10^6$ RMs, the RM$^2$ S/N rises to $88$--$353$ per bin, and the RM term of Equation~\ref{eqn:sigma_lnB}, $\sigma_\xi/2\boldsymbol{\xi}$, drops to $0.001$--$0.006$, negligible compared to the DM term. The precision then saturates at $\sigma_B/B=0.25$--$0.62$, set by the current uncertainty on gas distribution. The future FRB samples will remove this limitation. Combined with $3.6 \times 10^6$ RMs, $10^3$ FRBs yield $\sigma_B/B=0.09$--$0.41$, a factor of $2$--$4$ below the current value, and $10^4$ FRBs yield $0.03$--$0.13$, a factor of $6$--$13$ below the present precision. RMs and FRBs are therefore complementary: RMs provide the S/N, and FRBs provide the electron column density that converts RM into $B_\parallel$.

We note the scope of these forecasts. They apply to the BGS groups at $R<R_{200c}$, where the halo dominates the cross-correlation signal, and they do not include the RASS clusters, for which a cluster-specific forecast of the FRB cross-correlation is required. They refer to the statistical precision on $B_{\parallel,\mathrm{eff}}$, and thus do not include the uncertainty on the coherence length or the depolarization selection effect discussed in Section~\ref{subsec:b_estimator}. Finally, the precision on $\beta$ will in addition be limited by the uncertainty on $P_\mathrm{th}$ (Section~\ref{subsec:data_pth}), which is not reduced by the RM or FRB samples considered here.

\section{Discussion and Conclusion} \label{sec:discussion_and_conclusion}

In this work, we combined stacks of RM$^2$ around DESI BGS galaxy groups and RASS galaxy clusters with independent constraints on the electron column density and thermal pressure to derive limits on the line-of-sight magnetic field and the plasma $\beta$ in halos. The stacks are detected at total S/N of $1.9$ and $3.6$ respectively (Section~\ref{subsec:bfield_estimator}). For a coherence length $l_c=10$~kpc, the 95\% upper limits on the rms line-of-sight field are $< 3.9$--$11.6~\mu$G in groups and $< 2.3$--$6.4~\mu$G in clusters, and the 95\% lower limits on $\beta$ are $>0.31$ in groups and $>11$ in clusters (Figure~\ref{fig:B_beta_limits}), with the limits scaling as $l_c^{-1/2}$ and $l_c$, respectively. These limits imply that for a radially constant $\beta$, the magnetic pressure in clusters is at most $\approx10\%$ of the thermal pressure for $l_c=10$~kpc ($\beta>11$), a bound that scales as $l_c^{-1}$ (Figure~\ref{fig:B_beta_limits}, right), and reaches the $\beta\simeq100$ found in massive IllustrisTNG halos at $l_c\approx90$~kpc. In groups, the limits are weaker and still allow magnetic pressure comparable to the thermal pressure for $l_c=10$~kpc ($\beta>0.31$), so the current group data do not yet test field strengths in simulations. The field limits of $\lesssim2$--$12~\mu$G are far above primordial seed expectations, and thus probe the field amplified and redistributed by galaxy formation. These are one-sided bounds conditional on $l_c$ rather than measurements of the physical field: although the RM$^2$ stack itself is detected for the RASS clusters ($3.6\sigma$), its conversion to a field strength is subject to the limitations below.

First, the data constrain only combinations of the field strength and the coherence length (Equations~\ref{eqn:Brms} and~\ref{eqn:xi_model}), and the coherence length itself is not measured. Second, selection effects of polarized sources, most notably Faraday depolarization, can bias RM$^2$ low, and the limits then refer to the RM$^2$ amplitude that survives in the polarized-source catalog rather than to the physical field. The tighter limits for clusters than for groups at fixed $l_c$ should therefore not be interpreted as evidence for weaker fields in clusters. Third, we model each sample with a halo of its mean mass, rather than integrating over the halo mass distribution of the sample. Since RM$^2$ scales as the product of $n_\mathrm{e}^2$ and the squared field, this is an approximation. Finally, we assumed an isotropic random field with a single coherence length, and truncated the sightlines at $2R_{200c}$.

These limitations can be addressed with upcoming surveys. As forecasted in Section~\ref{sec:forecast} for DSA, $>10^6$ RMs, combined with $10^3$ and $10^4$ localized FRBs, will improve the fractional precision on $B_{\parallel,\mathrm{eff}}$ in groups from $0.38$--$0.75$ today to $0.09$--$0.41$ and $0.03$--$0.13$, respectively (Figure~\ref{fig:DM_RM_forecast}). Measurements of the kSZ and tSZ effects with next-generation CMB experiments, such as the Simons Observatory, will improve the electron density and thermal pressure profiles, and thereby the precision on $\beta$. The coherence length too can be estimated directly with DSA observations. First, the $\approx 3''$ DSA beam corresponds to $\approx 12$\,kpc at $\bar{z} = 0.26$, five times smaller than the ASKAP beam, and so the RM structure function measured using extended polarized sources behind the BGS groups and RASS clusters probes field fluctuations on scales of $\gtrsim 10$\,kpc, with a precision of $\sim 1.6$\,rad\,m$^{-2}$ per beam at the $8\sigma_p$ threshold. Second, spatial RM fluctuations within the beam depolarize the emission as a function of frequency. Depolarization stacks sample the magnetic field on scales below the beam, complementing the RM$^2$ stack, which is sensitive to $l_c$ in combination with the field strength.

\begin{figure*}[ht!]
\centering
\includegraphics[width=\textwidth]{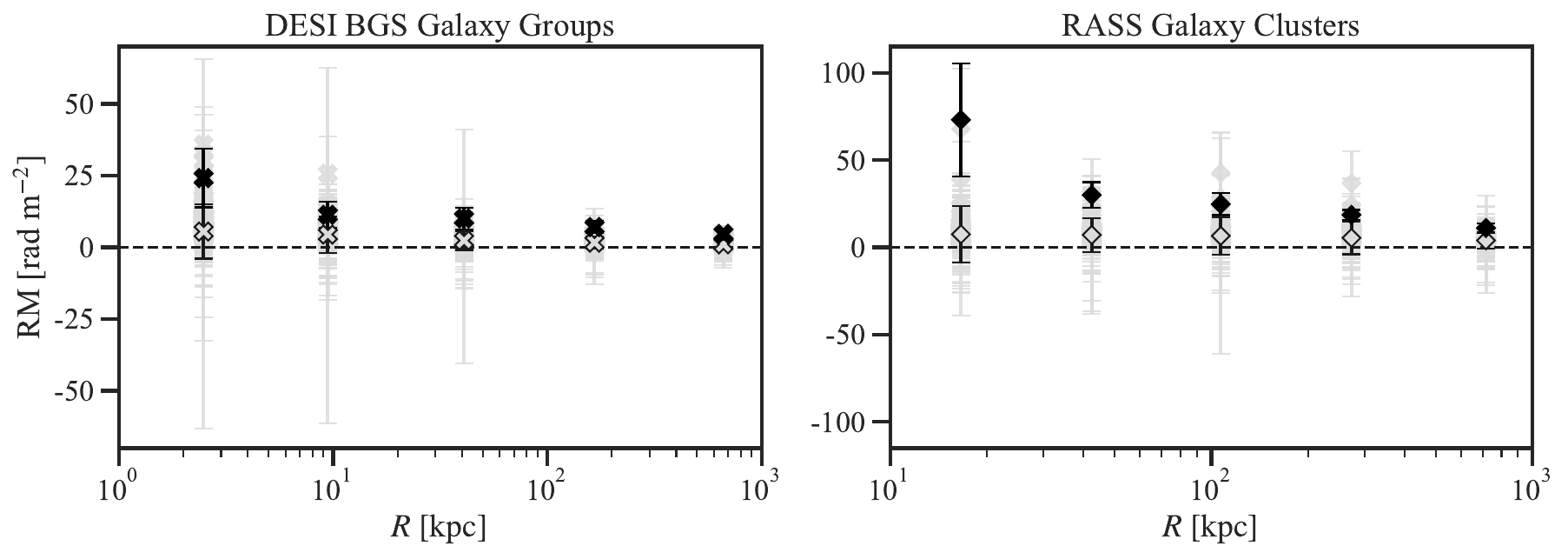}
\caption{Null-test validation of the correlation functions (in observer-frame) between SPICE-RACS Faraday rotation measures, and BGS galaxy groups (\textit{left}) and RASS galaxy clusters (\textit{right}). The covariance is estimated using jackknife resampling across 100 spatial patches. For the null tests, the Faraday rotation measures are randomly shuffled, with the resulting measurements from all 100 realizations shown as gray points. Open markers indicate the mean across the shuffled realizations, with uncertainties accounting for both shot noise and the uncertainty in the estimated mean. Filled markers indicates our measurement.}
\label{fig:null_test}
\end{figure*}

Beyond the stacked statistics presented here, a DSA-scale RM grid enables several other probes of cosmic magnetism. First, it makes individual systems accessible. For the grid of Section~\ref{subsec:dsa_rm}, a cluster at $z\approx0.05$ has $\sim130$ RMs within $R_{200c}$ and a group at $z\approx0.02$ has $\sim150$, uncertain by a factor of $\sim2$ given the range of source densities, with the $3''$ beam resolving $\sim1$--$3$\,kpc. This is sufficient to map the RM profiles, azimuthal asymmetries and structure functions of single halos, and to connect them with merger shocks and substructure~\citep{2021PASA...38...20A}. Second, the grid probes beyond the virial radius, allowing us to constrain magnetic fields in cosmic web filaments~\citep{2022MNRAS.512..945C, 2023MNRAS.518.2273C, 2024MNRAS.533.4068A}. Third, RM excesses around nearby galaxies, currently seen with $\sim2.5\times10^3$ LOFAR RMs~\citep{2023A&A...670L..23H}, can be mapped as a function of impact parameter and orientation to trace how galactic winds magnetize the circumgalactic medium. Finally, cross-correlating RM$^2$ with DESI tracers at higher redshift would trace the redshift evolution of halo magnetization~\citep{2026JCAP...07..096Z}.

In conclusion, RM$^2$ stacks around group- and cluster-mass halos, combined with independent measurements of the electron column density and thermal pressure, provide limits on the magnetic field strength and plasma $\beta$. The two main obstacles to turning these limits into measurements, the uncertain electron column density and the S/N of RM$^2$ stacks, will be largely removed by DSA-scale FRB and RM samples, which can thus break the baryon-magnetism degeneracy. Understanding the depolarization of RM stacks, and measuring the coherence length of the field, will be necessary to translate them into measurements of magnetic pressure support in halos, constrain galaxy formation models, assess its contribution to the tSZ deficit, and enable constraints on magnetogenesis scenarios.

\section*{Acknowledgments}
This material is based upon work supported in part by the National Science Foundation under CAREER Grant Number 2240032. This research is enabled in part by development funded by Schmidt Sciences for the Deep Synoptic Array project (DSA). The DSA is part of the Eric and Wendy Schmidt Observatory System in partnership with the California Institute of Technology. 
The IllustrisTNG simulations were undertaken with compute time awarded by the Gauss Centre for Supercomputing (GCS) under GCS Large-Scale Projects GCS-ILLU and GCS-DWAR on the GCS share of the supercomputer Hazel Hen at the High Performance Computing Center Stuttgart (HLRS), as well as on the machines of the Max Planck Computing and Data Facility (MPCDF) in Garching, Germany.

\facilities{Atacama Cosmology Telescope, Australian Square Kilometer Array Pathfinder, Dark Energy Spectroscopic Instrument, DES, DESI Legacy Imaging Surveys, CHIME, DSA-110, MeerKAT, Planck, RealFast, ROSAT, VLBI.}

\software{
\textsc{Astropy}~\citep{astropy:2013, astropy:2018, astropy:2022},
\textsc{BaryonForge}~\citep{2024OJAp....7E.108A}, 
\textsc{Cartopy}~\citep{Cartopy},
\textsc{Pyccl}~\citep{2019ApJS..242....2C},
\textsc{SciPy}~\citep{2020SciPy-NMeth},
\textsc{TreeCorr}~\citep{2015ascl.soft08007J}.
}

\appendix

\section{Null Tests}
\label{app:null_tests}

\begin{figure*}[ht!]
\centering
\includegraphics[width=\textwidth]{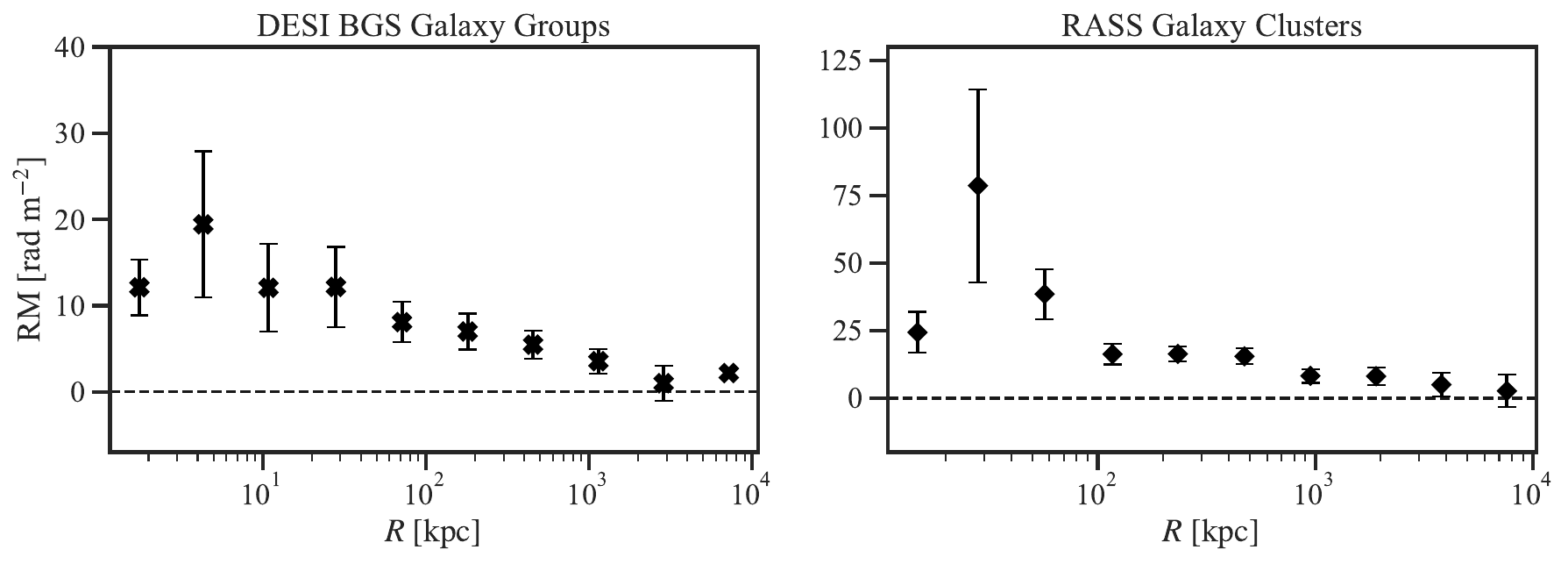}
\caption{RM amplitude, $\sqrt{\boldsymbol{\xi}}$ (observer frame), as a function of projected radius $R$ out to $\approx10$~Mpc, for DESI BGS galaxy groups (\textit{left}) and RASS galaxy clusters (\textit{right}), in 10 logarithmic bins, with $1\sigma$ jackknife uncertainties. The dashed line marks zero. The signal declines with radius and approaches zero at $\gtrsim$Mpc scales.}
\label{fig:null_test_large_scales}
\end{figure*}

To test whether the RM$^2$ cross-correlation measurements are affected by systematics unrelated to the association of RMs with halos, such as the survey geometry, variations in the source density, foreground residuals, or the estimator itself, we repeat the measurements after randomly shuffling the RMs among the polarized sources. Shuffling preserves the positions of the sources and the distribution of $\mathrm{RRM}^2$ values, but removes any physical correlation between $\mathrm{RRM}^2$ and the tracer positions. The expected cross-correlation is therefore zero in every radial bin. For each tracer sample, we generate 100 shuffled realizations, and measure $\boldsymbol{\xi}$ in the same radial bins and with the same pipeline and jackknife covariance as for the data (Section~\ref{subsec:bfield_estimator}). Figure~\ref{fig:null_test} shows the individual realizations as gray points, and their mean as open markers, with uncertainties that account for both the shot noise and the uncertainty in the estimated mean. The mean of the shuffled measurements is consistent with zero in all radial bins for both samples. We therefore find no evidence that the stacks of Section~\ref{subsec:bfield_estimator} are produced by the survey geometry or any residual correlated signal.

\section{RM Profiles at Large Projected Radii}
\label{app:large_radii}

The stacks in the main text extend to $R=1$~Mpc. To test whether the RM$^2$ signal is localized around the tracers, Figure~\ref{fig:null_test_large_scales} extends the measurement to $R\approx10$~Mpc, with 10 logarithmic bins and the same pipeline and jackknife covariance as in Section~\ref{subsec:bfield_estimator}. For both samples, the RM amplitude declines monotonically with radius, from $\approx10$--$20$~rad\,m$^{-2}$ (BGS) and $\approx25$--$80$~rad\,m$^{-2}$ (RASS) at $R\lesssim50$~kpc to a few rad\,m$^{-2}$ at $R\gtrsim3$~Mpc. At these radii, the RASS profile and the BGS bin at $R\approx1$~Mpc are consistent with zero within $\approx1$--$1.5\sigma$, noting that $\sqrt{\boldsymbol{\xi}}$ is positive by construction, so that a vanishing signal appears as a small positive RM.

The decline of the signal beyond the halo scale supports that the excess is associated with the tracers, rather than being a large-scale offset in $\mathrm{RRM}^2$ near the tracers relative to random positions, for example from residual Galactic foregrounds or the survey geometry, which would persist to large radii. Any such offset is limited to a few rad\,m$^{-2}$, well below the signal at small radii. This complements the null tests of Appendix~\ref{app:null_tests}: shuffling the RMs gives zero at every radius by construction, and thus validates the estimator, whereas the decline of the measured signal at large radii tests the large-scale behavior of the data themselves.

\bibliography{manuscript}{}
\bibliographystyle{aasjournal}

\end{document}